\documentclass[twocolumn,amsmath,amssymb,aps,prl,superscriptaddress,nofootinbib]{revtex4-2}

\usepackage[english]{babel}
\usepackage[utf8]{inputenc}
\usepackage[colorlinks]{hyperref}
\usepackage{xr-hyper}

\usepackage[T1]{fontenc}
\usepackage[dvips]{graphicx}
\usepackage{amsmath}
\usepackage{amsfonts}
\usepackage{color}
\usepackage{ulem}
\usepackage[caption=false, justification=justified]{subfig}
\usepackage{url}
\usepackage{ulem}
\usepackage[top=2cm, bottom=2cm, left=2cm, right=2cm]{geometry}

\makeatletter
\newcommand*{\addFileDependency}[1]{% argument=file name and extension
 \typeout{(#1)}
 \@addtofilelist{#1}
 \IfFileExists{#1}{}{\typeout{No file #1.}}
}
\makeatother

\newcommand*{\myexternaldocument}[1]{%
 \externaldocument{#1}%
 \addFileDependency{#1.tex}%
 \addFileDependency{#1.aux}%
}

\myexternaldocument{supplementary_information_PRL}

\usepackage{xcolor}

\hypersetup{filecolor=blue}
\hypersetup{citecolor=blue}
\hypersetup{urlcolor=blue} 

\hypersetup{linkcolor=blue}

\renewcommand{\selectlanguage}[1]{}

\begin{document}

\title{Interfacial spin-orbitronic effects controlled with different oxidation levels at the Co|Al interface}

\author{Sachin~Krishnia}
\email{sachinbudana@gmail.com}
\thanks{Present address: Institute of Physics, Johannes Gutenberg University Mainz, Staudingerweg 7, 55128 Mainz, Germany}
\affiliation{Laboratoire Albert Fert, CNRS, Thales, Université Paris-Saclay, 91767, Palaiseau, France}
\author{Libor~Vojáček}
\affiliation{Université Grenoble Alpes, CEA, CNRS, IRIG-SPINTEC, 38000 Grenoble, France}
\author{Tristan~Da~Câmara~Santa~Clara~Gomes}
\thanks{Present address: INESC Microsistemas e Nanotecnologias (INESC MN) Rua Alves Redol, 9 , 1000-029 Lisbon, Portugal}
\affiliation{Laboratoire Albert Fert, CNRS, Thales, Université Paris-Saclay, 91767, Palaiseau, France}
\author{Nicolas~Sebe}
\affiliation{Laboratoire Albert Fert, CNRS, Thales, Université Paris-Saclay, 91767, Palaiseau, France}
\author{Fatima~Ibrahim}
\affiliation{Université Grenoble Alpes, CEA, CNRS, IRIG-SPINTEC, 38000 Grenoble, France}
\author{Jing~Li}
\affiliation{Université Grenoble Alpes, CEA, Leti, F-38000 Grenoble, France}

%\author{Yanis Sassi}
%\affiliation{Laboratoire Albert Fert, CNRS, Thales, Université Paris-Saclay, 91767, Palaiseau, France}
\author{Luis~Moreno~Vicente-Arche}
\affiliation{Laboratoire Albert Fert, CNRS, Thales, Université Paris-Saclay, 91767, Palaiseau, France}
\author{Sophie~Collin}
\affiliation{Laboratoire Albert Fert, CNRS, Thales, Université Paris-Saclay, 91767, Palaiseau, France}
\author{Thibaud~Denneulin}
\affiliation{Ernst Ruska-Centre for Microscopy and Spectroscopy with Electrons (ER-C-1), Forschungszentrum Jülich GmbH, 52425 Jülich, Germany}
%\author{András Kovács}
%\affiliation{Ernst Ruska-Centre for Microscopy and Spectroscopy with Electrons (ER-C 1) and Peter Grünberg Institut (PGI-5), Forschungszentrum Jülich GmbH, 52425 Jülich, Germany}
\author{Rafal E.~Dunin-Borkowski}
\affiliation{Ernst Ruska-Centre for Microscopy and Spectroscopy with Electrons (ER-C-1), Forschungszentrum Jülich GmbH, 52425 Jülich, Germany}
\author{Philippe~Ohresser}
\affiliation{Synchrotron SOLEIL, L’Orme des Merisiers, 91190, Saint Aubin, France}
\author{Nicolas~Jaouen}
\affiliation{Synchrotron SOLEIL, L’Orme des Merisiers, 91190, Saint Aubin, France}
%\author{Jean-Marie George}
%\affiliation{Unité Mixte de Physique, CNRS, Thales, Université Paris-Saclay, 91767, Palaiseau, France}
\author{André~Thiaville}
\affiliation{Laboratoire de Physique des Solides, Université Paris-Saclay, CNRS, Orsay 91405, France}
\author{Albert~Fert}
\affiliation{Laboratoire Albert Fert, CNRS, Thales, Université Paris-Saclay, 91767, Palaiseau, France}
\author{Henri~Jaffrès}
\affiliation{Laboratoire Albert Fert, CNRS, Thales, Université Paris-Saclay, 91767, Palaiseau, France}
\author{Mairbek~Chshiev}
\affiliation{Université Grenoble Alpes, CEA, CNRS, IRIG-SPINTEC, 38000 Grenoble, France}
\affiliation{Institut Universitaire de France, 75231, Paris, France}
\author{Nicolas~Reyren}
\affiliation{Laboratoire Albert Fert, CNRS, Thales, Université Paris-Saclay, 91767, Palaiseau, France}
\author{Vincent~Cros}
\email{vincent.cros@cnrs-thales.fr}
\affiliation{Laboratoire Albert Fert, CNRS, Thales, Université Paris-Saclay, 91767, Palaiseau, France}

\begin{abstract}
\textbf{Abstract:} 
Perpendicular magnetic anisotropy (PMA) and Dzyaloshinskii-Moriya interactions are key interactions in modern spintronics. These interactions are thought to be dominated by the oxidation of the Co|Al interface in the archetypal Platinum-Cobalt-Aluminum oxide system. Here, we observe a double sign change in the anisotropy and about threefold variation in interfacial chiral interaction, influenced not only by the oxidation, but also by the metallic Al thickness. Contrary to previous assumptions about negligible spin-orbit effects at light metal interfaces, we not only observe strong PMA with fully oxidized Al, decreasing and turning negative (in-plane) with less oxygen at the Co|Al interface, we also observe that the magnetic anisotropy reverts to positive (out-of-plane) values at fully metallic Co|Al interface.  These findings suggest modification in Co \textit{d}-band via Co|Al orbital hybridization, an effect supported by X-ray absorption spectroscopy and  \textit{ab initio} theory calculations, highlighting the key impact of strain on interfacial mechanisms at fully metallic Co|Al interface.

\end{abstract}

\maketitle

Breaking of inversion symmetry at interfaces together with the presence of a sizeable spin-orbit coupling (SOC) are the two major components of a collection of phenomena in modern  surface magnetism and spintronics~\cite{SoumyanarayananNatReview2016}. These are interfacial effects such as the magnetic anisotropy~\cite{Dieny2017}, the interfacial Dzyaloshinskii–Moriya interaction (i-DMI)~\cite{Thiaville_2012,Fert2023JPS} as well as charge-to-spin and charge to orbit interconversion mechanisms~\cite{Miron2011,Krishnia2023}. Early experiments in the 90's considered bilayers of ferromagnet and heavy material films (or multilayers), an archetypal structure being Pt|Co bilayers, to exhibit large interfacial perpendicular magnetic anisotropy (PMA). In these systems, the large SOC of Pt combined with the orbital hybridization at Pt|Co interface results in strong interfacial PMA~\cite{Nakajima1998,CarciaAPL,WellerPRB}. Moreover, triggered by the search of large room temperature tunnel magnetoresistance effect in the 2000's %~\cite{Moodera1995}
, a large number of studies ~\cite{Rodmacq2003,Monso2002,Manchon2008} have been then carried out on multilayered systems composed of a heavy material e.g., Pt, a transition metal magnetic material e.g., Co, and an oxide e.g., Al$_2$O$_3$ or MgO showing large effective PMA. Pioneer experiments on Pt|Co|AlO$_x$ trilayers highlighted the necessity of optimum interface oxidation and established a paradigm for obtaining effective PMA~\cite{Monso2002,Manchon2008,Dieny2017,Balan2024}. In particular, it was shown that the effective PMA drastically diminishes with over or under oxidation of Co|AlO$_x$ interface with a magnetization of Co that can be found in some cases to be in-plane even for 0.6\,nm-thick Co films. Remarkably, in these experiments and in follow-up theoretical work, the interfacial anisotropy at the transition metal|oxide interfaces was found to be of similar amplitude to that of the Co|Pt interface, despite the involvement of light elements with weak spin-orbit coupling, namely Co, Al and O ~\cite{Dieny2017,Monso2002}. The origin of large PMA in optimum Pt|Co|AlO$_x$ structures has been associated with the hybridization between Co 3\textit{d} and O 2\textit{p} orbitals similar to the case of Co|MgO interfaces~\cite{Manchon2008,Yang2011,Peng2015}. However, the underlying physical mechanism of PMA to in-plane effective anisotropy transition with under oxidation of Co|AlO$_x$ interface is still under debate after two decades. The recent prediction and observation of enhanced PMA at Co|graphene and Co|hexagonal Boron Nitride interfaces  further challenges conventional understanding of the origin of these effects~\cite{Rougemaille2012,Yang2016,Yang2018,Hallal2021,El-Kerdi2023}.   Recent results show that, with Al protected from oxidation, a purely metallic Co|Al  interface exhibits an orbital texture which gives rise to efficient charge to orbit interconversion and very large enhancement of the current-induced torques on Co~\cite{Krishnia2023} and is confirmed  by \textit{ab-initio} calculations of orbital accumulation~\cite{Nikolaev}. The present study aims to demonstrate that the richness of these interfaces is still surprisingly far from having been fully exploited, both from a physics point of view and for improving the performance of spintronic devices, offering a new paradigm of spin-orbit phenomena in them.

\medskip

Another crucial interaction at play at the interface between a magnetic film and a light metal or an oxide film is the interfacial DMI that promotes the development of chiral magnetism and the associated different types of chiral spin textures e.g. chiral domain walls, skyrmions etc.
%~\cite{Fert2013,Jiang2015,Woo2016,Moreau-Luchaire2016,Boulle2016,Litzius2017}
Of interest for the present study, it has been shown that the Pt|Co|AlO$_x$ interface ~\cite{Belmeguenai2015} exhibits one of the largest reported i-DMI. However, the values measured on this system by various groups in the literature range from  0.5 to 2.5 pJ/m~\cite{Kim2017,Conte2017,Xin2018,Kuepferling2021}. Therefore, it is crucial to investigate how the i-DMI evolves with the oxidation degree of Al close to the interface and its microscopic origin in Pt|Co|(Al|)AlO$_x$ multilayer. In this scenario, general questions arise about the physical mechanism through which the orbital hybridization between Al 3\textit{p} and Co 3\textit{d} orbits impacts the effective PMA and DMI, and about the need to oxidize the Co|AlO$_x$ interface to maximize the effects, which have been seen as long-established requirement for more than two decades and have not been completely solved yet.
\medskip

\medskip

In this Letter, we thoroughly investigate the origin of several of these interfacial effects related to breaking of the inversion symmetry, orbital hybridization and spin-orbit interaction at the model interface system, namely Pt|Co|Al*, where Al* refers to the natural oxidation of Al in air. By varying the deposited Al thickness, the Co|Al* interface termination is controlled to vary between Co-O-Al interface bonds, that is an FM|oxide interface and Co-Al interface bonds, that corresponds to FM|metal interface. The most striking result of our study is that the amplitude of the magnetic anisotropy strongly varies depending on the exact constitution of the interface. Starting with a large perpendicular magnetic effective anisotropy at Co(0.9\,nm)|AlO$_x$ interface (in agreement with previous works), it then rapidly decreases with increasing metallic Al and even undergoes a sign change. Importantly, we report here that it experiences another sign change, overlooked until now, for thicker nominal Al, before it saturates upon the formation of a thicker metallic interface with stronger PMA. Such double sign change of effective anisotropy at the weak SOC Co|Al* interface suggests a mechanism involving orbital hybridization that we verify by performing X-ray absorption spectroscopy (XAS) measurements combined with density functional theory (DFT) calculations. In a correlated way, we find that the net i-DMI decreases with reduced oxygen atomic fraction at the Co|Al* interface, ultimately saturating for a fully metallic Co|Al interface, at a level of about one third of the fully oxidized Co(0.9\,nm)|Al* interface. It demonstrates again the strength of the DMI at the metallic Co|Al interface ~\cite{Ajejas2022}, which we find to be of similar magnitude as the one at the Pt|Co. These experimental results qualitatively match first principles calculations in which the observed behavior is correlated with the interfacial dipole strength.

\medskip

\begin{figure}
 \centering
 \includegraphics[width=0.47\textwidth]{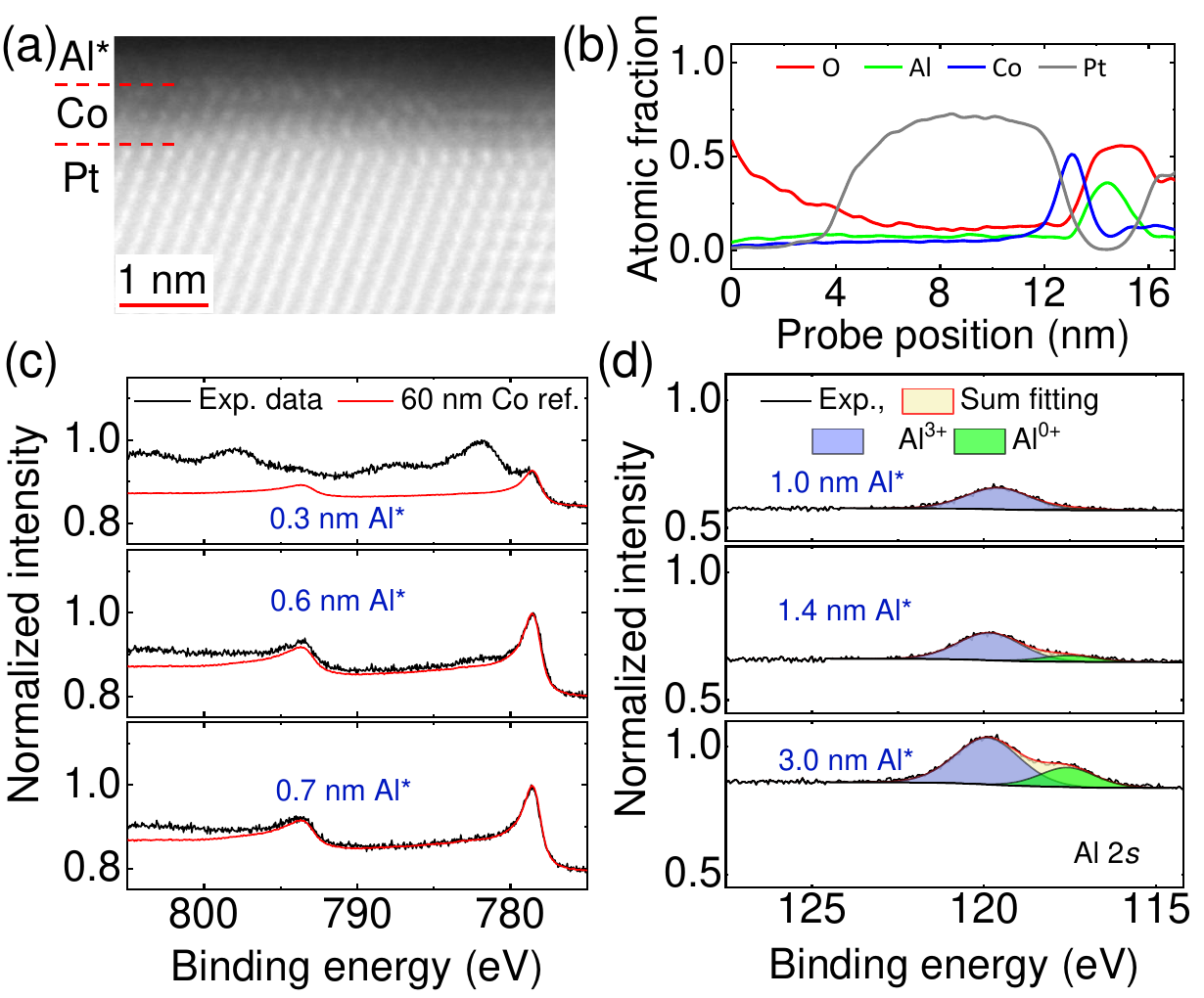}
\caption{Structural and chemical characterization of the multilayer. (a) HAADF STEM image showing the contrast of Pt|Co|Al*(0.7) sample. (b) Depth-profile of integrated (over 30\,nm) atomic fraction of Pt (light gray), Co (blue), Al (green) and O(red) along the thickness. %(c) Structure of bilayer Co|(O$_\rm -x$)$\rm Al_2 \rm O_3$ used in the \textit{ab-initio} calculations. 100$\%$ oxidized interface is shown between the 4-monolayer hcp Co slab and $\alpha -\rm Al_2 \rm O_3$ with pseudo-hydrogen passivation. 
(c) XPS spectra (black) near the Co 2\textit{p} energy from Pt|Co|Al*($t_{\rm Al}*$) for $t_{\rm Al*}$ = 0.3, 0.6 and 0.7 nm, with a reference spectrum of in-situ grown 60 nm Co (red). (d) XPS spectra (black) near Al 2\textit{s} energy from Pt|Co|Al*($t_{\rm Al}*$) for $t_{\rm Al}*$ = 1.0, 1.4 and 3 nm. Fitted peaks of Al$^{3+}$ and Al$^{0}$ are shown by blue and green envelopes, while the total areas are filled with light yellow.}
\label{Fig1}
\end{figure}

\medskip

Our approach is to fine-tune the oxygen content at the Co|Al* interface in series of samples of composition Si(sub.)|SiO$_2$(280)|Ta(5)|Pt(8)|Co(0.9)|Al*($t_{\rm Al*}$), where the numbers in parentheses are the nominal thicknesses in nanometers. The absolute error on the thickness is evaluated to be below 0.1\,nm, corresponding to the symbol size in the following figures. The samples are grown by dc magnetron sputtering onto thermally oxidized Si|SiO$_2$ (280 nm) substrates. The 5-nm thick Ta seed layer is introduced to ameliorate the adhesion and the surface roughness of the consecutive layers.  It also results in an increase of PMA at the Pt|Co interface by improving the Pt texture~\cite{Parakkat_2016}. The bilayer composed of 8\,nm of Pt and 0.9\,nm of Co is kept nominally the same in all the series. We only modulate the top Al layer with a systematic variation in Al nominal thickness in steps of 0.1\,nm. Note that it is well known that the bottom Pt|Co interface is the site of several interfacial effects related to inversion symmetry breaking and SOC, such as a large PMA identified in the 1990s~\cite{Carcia1990}, and more recently, strong i-DMI observed since 2015~\cite{Belmeguenai2015}. The main objective of this study is to investigate how the total effective magnetic anisotropy and the i-DMI are modified by adding, on top of Co, a film of a light element, namely Al, with different oxidation states on top of this model Pt|Co bilayer. 

\medskip

To this aim, it is crucial to distinguish intrinsic effects related to electronic and orbital effects (such as those calculated {\it ab initio} for instance) from extrinsic structural effects, such as intermixing, grain structure, or strain. Prior to  focusing on the evolution of the anisotropy and i-DMI, we hence investigate both the chemical and structural nature of our samples, with a particular focus on the interface quality. 

\medskip

We first employ the scanning transmission electron microscopy (STEM) technique. In Fig.~\ref{Fig1}a, we present a high-angle annular dark field (HAADF) image of a Pt|Co|Al*(0.7) sample together with the spatial distribution of the different elements, namely Pt (light gray), Co (blue), Al (green) and O (red) obtained using energy dispersive X-ray spectroscopy (EDX) technique (see Fig.~\ref{Fig1}b). The STEM and EDX measurements demonstrate a very homogeneous polycrystalline growth of the different layers in the stack as well as the existence of sharp interfaces both between Pt|Co and Co|Al* (see SI for more details\cite{supmat}).

\medskip

To access more precisely the oxidation state of Co and Al atoms in Pt|Co|Al*($t_{\rm Al*}$) series, we perform X-ray photoelectron spectroscopy (XPS) experiments on samples with various $t_{\rm Al*}$. In Fig.~\ref{Fig1}c, we present the XPS spectra (black curve) collected at Co 2\textit{p} levels for three different thickness of Al*, namely 0.3, 0.6 and 0.7\,nm. In addition, we also include in Fig.~\ref{Fig1}c (red curve), the XPS spectrum measured on an in-situ grown 60-nm thick Co film considered as a reference for metallic Co. We clearly see that the spectrum for $t_{\rm Al*} = 0.3$\,nm is very different from the reference Co, indicating the partial oxidation of Co with two peaks at 778.6\,eV and 781.9\,eV correspond to CoO 2$p_{3/2}$ and Co 2$p_{3/2}$, respectively. The CoO 2$p_{3/2}$ peak at 781.9\,eV is strongly decreased for $t_{\rm Al}* = 0.6$\,nm, and has even completely disappeared for $t_{\rm Al}* = 0.7$\,nm. In consequence, we conclude from these XPS characterizations that 0.7\,nm of Al* is thick enough to completely protect Co from oxidation (at least over the several months of the experiments). Thus in the following, we will not consider samples with an Al* thickness thinner than this threshold value. An important complementary information is to determine the presence of either metallic or oxidized Al at the Co|Al* interface. Therefore, we also measured the XPS spectra at the Al 2$s$ edge for several $t_{\rm Al*}$ as shown in Fig.~\ref{Fig1}d. The spectra obtained for $t_{\rm Al*}\leq 1$\,nm reveal solely the Al$^{3+}$ valence state, consistent with a full oxidization of the Al film. Above 1\,nm, the Co|Al* interface start to display a metallic character, as demonstrated by the presence of a peak associated with metallic Al (Al$^{0}$). Finally, we emphasize that the conclusions obtained from the analysis of the XPS measurements are quantitatively consistent with the ones from X-ray absorption spectroscopy (XAS) shown in Supplementary Information (SI)~\cite{supmat}.

\medskip

\begin{figure}
 \centering
 \includegraphics[width=0.47\textwidth]{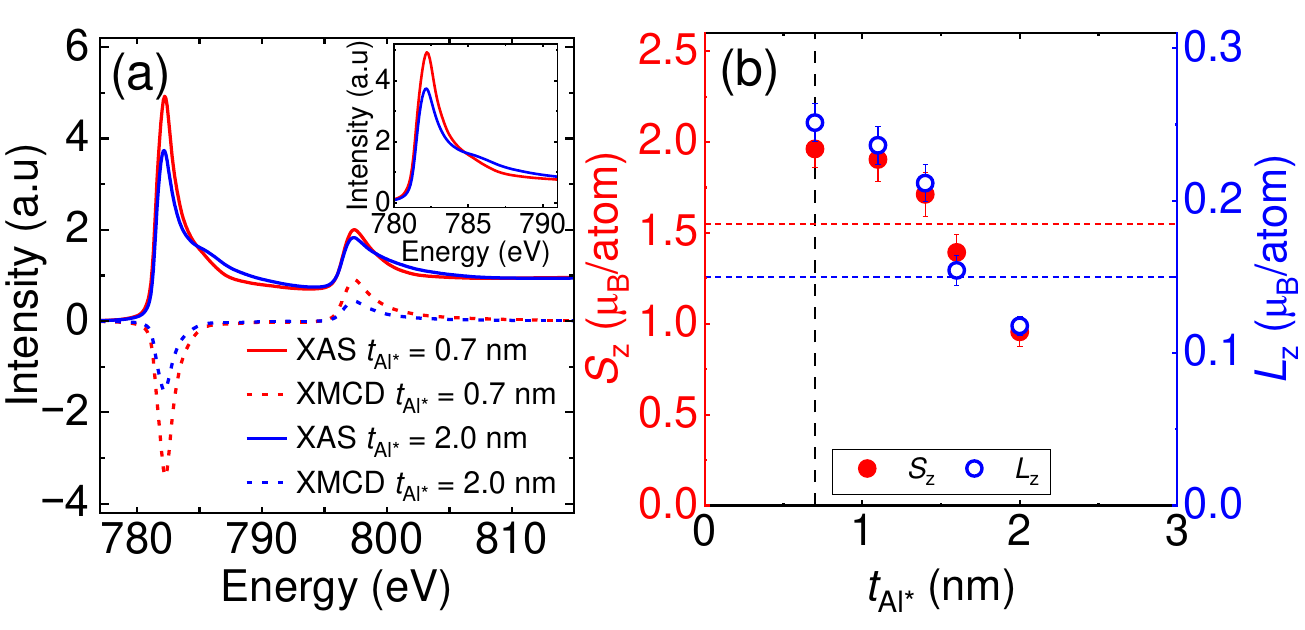}
\caption{(a) XAS and XMCD spectra from the Co $L_3$ and $L_2$ edges in Pt|Co|Al* for $t_{\rm Al*} = 0.7$\,nm (red) and 2.0\,nm (blue) at room temperature. The inset shows a zoom of XAS spectra around the $L_3$ edge. (b) Spin and orbital magnetic moments of Co atoms as a function of Al* thickness estimated from sum rules. The vertical dotted lines corresponds to $t_{\rm Al*}$ = 0.7 nm, that is the minimum Co thickness for ensuring non-oxidized Co. The two horizontal lines indicate the spin and orbital moments for bulk Co (from \cite{CTChenPRL1995}). 
 }
\label{Fig2}
\end{figure}

Next, we quantitatively determine the spin ($S_z$) and orbital ($L_z$) moments of Co using x-ray absorption (XAS) and x-ray magnetic circular dichroism (XMCD) at the Co $L_{2,3}$ edges. In Fig.~\ref{Fig2}a, we display the mean XAS and the XMCD spectra for two samples with different $t_{\rm Al*}$ thickness, namely 0.7 and 2\,nm. For 0.7\,nm, the case of absence of Co oxidation and full oxidization of Al film according to XPS, the shapes of both XAS and XMCD spectra are typical of those recorded for bulk Co \cite{CTChenPRL1995}. Applying the XMCD sum rules, we find $S_z = 1.8\,\mu_\mathrm{B}$/atom that is by about 15\% higher than the value for  bulk Co. For $t_{\rm Al*} = 2$\,nm, the existence of Co-Al bonds associated with the presence of a non-oxidized Al film result in some noticeable changes of the XAS spectra at both $L_3$ and $L_2$ edges (Fig.~\ref{Fig2}a), notably with a strong drop of peak intensity associated to a large variation of the number of unoccupied Co $d$-states as well as the appearance of a shoulder peak after the $L_3$ edge. This latter reflects a significant change in the Co band structure in the case of the Co|Al interface. In Fig. \ref{Fig2}b, we present how the transition between Co-O-Al bonds to Co-Al impacts both the evolution of spin ($S_z$) and orbital ($L_z$) Co magnetic moments when $t_{\rm Al*}$ increases. Both $S_z$ and $L_z$ remain slightly higher than bulk Co values up to $t_{\rm Al*} = 1.4$\,nm before they both continuously drop, reaching an $S_z$ value of $1\,\mu_\mathrm{B}$/atom for $t_{\rm Al*} = 2$\,nm. The observed evolution of the magnetic moments with Al* thickness has been confirmed by SQUID magnetometry (see SI~\cite{supmat}). 

\medskip

To understand the microscopic origin of the evolution of Co spin and orbital moments with $t_{\rm Al*}$, but also as described later, of i-DMI and the anisotropy, we have performed \textit{ab initio} calculations within DFT/PBE~\cite{wang_correlation_1991,perdew_generalized_1996-2} as implemented in \texttt{VASP}~\cite{kresse_ab_1993-1,kresse_efficiency_1996-1, kresse_efficient_1996, kresse_ultrasoft_1999}. For that, we have constructed a heterostructure of a four-monolayer hcp(0001) Co film and $\alpha$-Al$_2$O$_3$ with varying interface oxidation and Al insertion in-between. See the Supplemental Material~\cite{supmat} for details on the DFT methodology, the simulated structures, and the calculations of spin and orbital magnetic moments. The experimental results shown in Fig. \ref{Fig2}, are found to be in good quantitative agreement with the DFT ones. First, we note that the large $S_z$ for $t_{\rm Al*} = 0.7$\,nm is attributed mainly to a higher $S_z$ for the Co atoms at the interface with the oxide, similarly to what was found at Co|MgO interface~\cite{Yang2011}. Second, we find that as the Co|Al* interface becomes metallic with increasing Al* thickness, the interfacial Co acquires a negative charge, leading to a reduced magnetic moment of about~20\%.

\medskip

\begin{figure}
 \centering
 \includegraphics[width=0.47\textwidth]{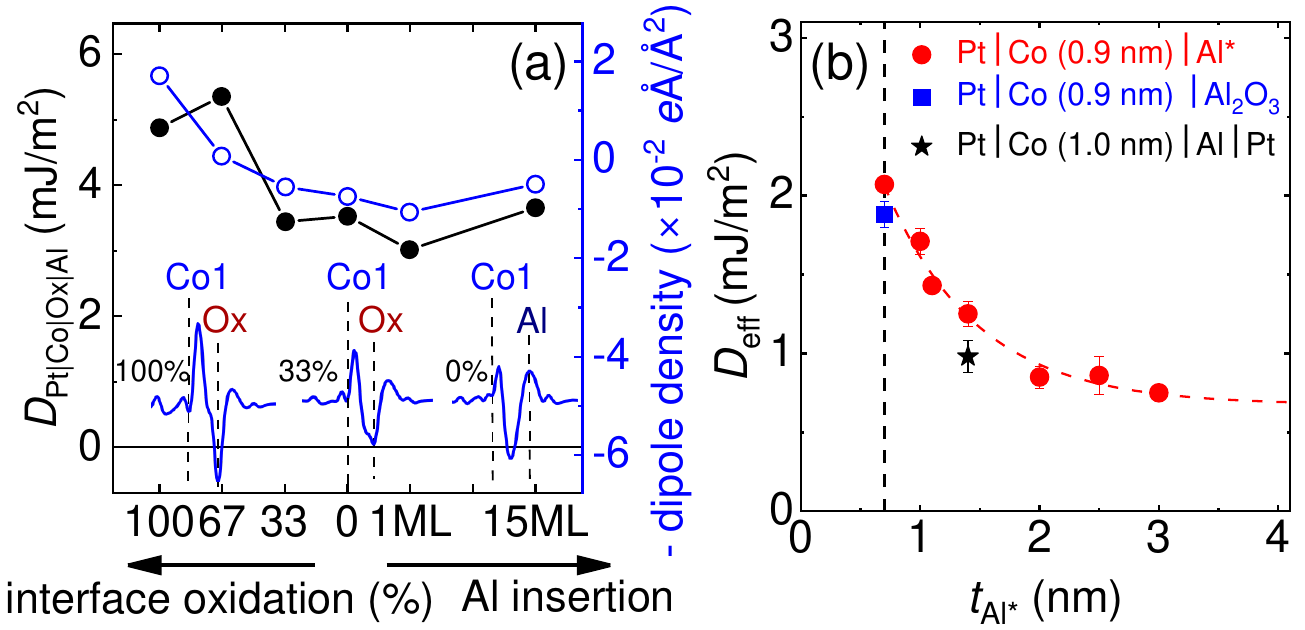}
\caption{DMI modulation with the oxidation level at the Co|Al* interface. (a)~Effective DMI coefficient of Pt|Co|(O$\rm x$)$\rm Al_2 \rm O_3$ calculated from \textit{ab-initio} total-energy method. Its decrease is correlated to the Co|Ox interfacial dipole (Rashba-type DMI) formed by charge transfer to the oxygen as shown in insets for 100\%, 33\% and 0\% O content at the interface. (b)~Experimental variation in effective DMI measured by Brillouin light scattering, as a function of $t_{\rm Al*}$ (red). The DMI values from other studies in the case of oxide (Pt|Co|$\rm Al_2 \rm O_3$)~\cite{Belmeguenai2015} and metallic (Pt|Co|Al|Pt)~\cite{Ajejas2022} top interfaces are also displayed by a blue square and a black star, respectively. Assuming an i-DMI independent of the Co thickness, the Pt|Co(0.9)|Al point can be calculated (black star).
}
\label{Fig3}
\end{figure}
\medskip

\begin{figure}
 \centering
 \includegraphics[width=0.47\textwidth]{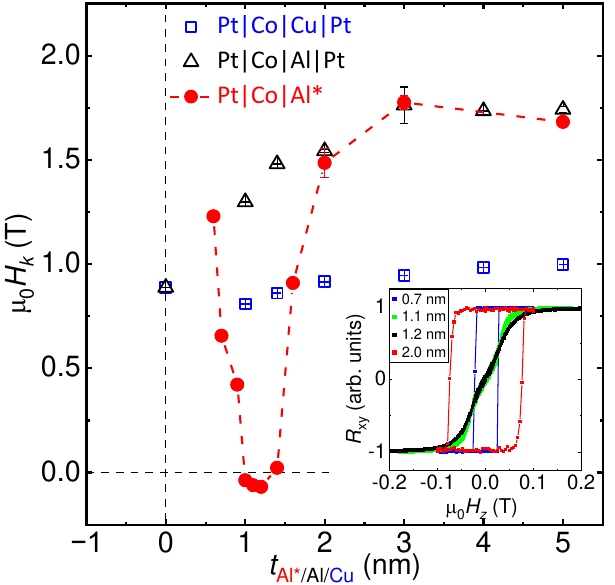}
\caption{Out-of-plane effective anisotropy field as a function of Al* (red), Al (black) and Cu (blue) thicknesses in Pt|Co|Al*($t_{\rm Al}*$), Pt|Co|Al|($t_{\rm Al}$)|Pt and Pt|Co|Cu($t_{\rm Cu}$)|Pt samples. Inset shows the normalized transverse resistance $R_{xy}$ as a function of the out-of-plane magnetic field in Pt|Co|Al* ($t_{\rm Al}*$) for $t_{\rm Al*} = {0.7, 1.1, 1.2, 2.0}$\,nm. 
}
\label{Fig4}
\end{figure}

In Fig.~\ref{Fig3}a, we present the DFT results for the evolution of the i-DMI constant that is obtained from the energy difference between the clockwise and anticlockwise spin configurations~\cite{Yang2011}. To compare with the experimental values, several Co|Al* structures have been  constructed in which the oxygen is gradually removed from the interface (see SI for the structures visualization~\cite{supmat}). In Fig.~\ref{Fig3}b, we display how the effective DMI $D_{\rm eff}$ amplitude measured by Brillouin light scattering (BLS) spectroscopy is evolving as a function of the Al* thickness. The large DMI value measured for $t_{\rm Al*}= 0.7$\,nm, close to 2\,$\mathrm{mJ}/\mathrm{m}^2$, is found to be closely matching the one measured by Belmeguenai {\it et al.}~\cite{Belmeguenai2015}, where they deposited an oxide layer on top of the Co, (Pt|Co|$\rm Al_2 \rm O_3$) (blue square in Fig.~\ref{Fig3}b). Then with increasing $t_{\rm Al*}$, $D_{\rm eff}$ exhibits an exponential-like decay, saturating for $t_{\rm Al*} \gtrsim 2$\,nm (decay length is about 1.3\,nm). For $t_{\rm Al*} = 1.4$\,nm, the measured DMI is very close to the fully metallic Al case when caped by Pt that was measured by BLS in Ta(5)|Pt(8)|Co(1.0)|Al(1.4)|Pt(3) in a previous study~\cite{Ajejas2022} (black star in Fig.~\ref{Fig3}b). Note that this last value is slightly smaller, but once the DMI value is renormalized by the Co thickness, assuming that the DMI can be written as $D_{\rm eff} = D_{\rm s}/t_{\rm Co}$ ($D_{\rm s}$ is i-DMI energy), we find the star, which coincides with our $t_{\rm Al*}=1.4$\,nm data. Given that the bottom Pt|Co interface is identical in all samples, and considering the weak expected SOC at the top Co|Al* interface, the experimental decrease in $D_{\rm eff}$ with $t_{\rm Al*}$ is attributed to the variation in the electronic filling of the Co \textit{d}-band at the Co|Al* interface, which depends on the thickness of $t_{\rm Al*}$, as revealed by our XAS and XMCD measurements (Fig.~\ref{Fig2}). Interestingly, we find that the observed trend is reproduced by the \textit{ab initio} calculations (black curve in Fig.~\ref{Fig3}a), with a decreasing tendency when oxygen is removed from the Co|Al$_2$O$_3$ interface. This behaviour is explained by the electronic transfer from the Cobalt to the Oxygen nearby atoms that is stronger for larger oxidation, and that creates an interfacial charge dipole (blue curve in Fig.~\ref{Fig3}a). This electrical dipole induces a Rashba-type DMI~\cite{belabbes_oxygen-enabled_2016,Yang2018SciReport} at the Co|Al* interface, which is additive to the DMI contribution from the bottom Pt|Co. See the Supplemental Material~\cite{supmat} for details of the BLS experiments and \textit{ab initio} DMI calculation.

\medskip

\begin{figure}
 \centering
 \includegraphics[width=0.47\textwidth]{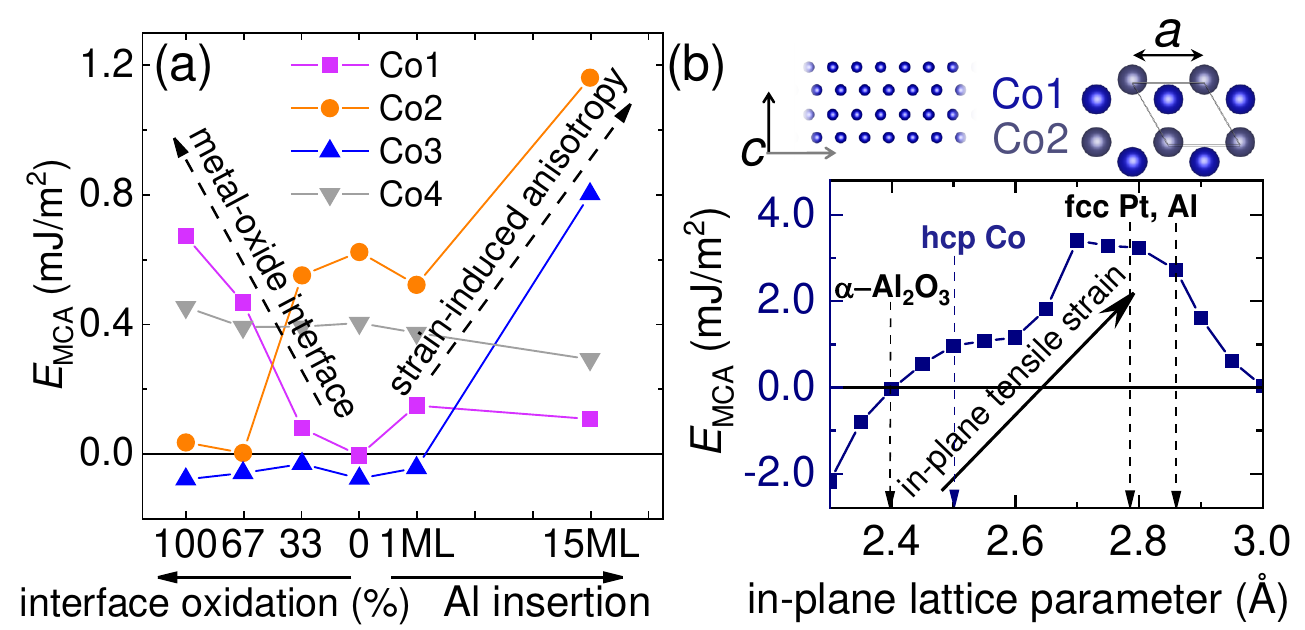}
\caption{Magnetocrystalline anisotropy energy calculated from {\it ab initio} total energies in Co layers (Co1 is at the interface) as a function of (a) interfacial oxygen content and (b) in-plane lattice parameter. The (vertical dashed lines in (b) indicate in-plane lattice parameters. 
}
\label{Fig5}
\end{figure}
\medskip

Another striking result of the present study is the strong and unexpected Al thickness dependence of the magnetic anisotropy in the Pt|Co|Al* system, which has been used about two decades ago for optimizing the perpendicular anisotropy ~\cite{Manchon2008,Dieny2017}. In Fig.~\ref{Fig4}, we present the experimental measurements of out-of-plane effective anisotropy field ($\mu_0H_K$) versus Al* thickness. The effective anisotropy has been determined through the anomalous Hall effect, measuring the transverse resistance ($R_{xy}$) as a function of in-plane and out-of-plane magnetic field,  on 5-$\mu$m wide Hall bars. In the inset of Fig.~\ref{Fig4} are plotted several normalized $R_{xy}$ curves as a function of out-of-plane magnetic field ($H_z$) in Pt|Co|Al*($t_{\rm Al}*$) samples for a few relevant $t_{\rm Al}*$. As demonstrated above by our XPS measurements (Fig.~\ref{Fig1}), the Co remains non-oxidized for $t_{\rm Al}* = 0.7$\,nm, resulting in a square hysteresis loop with sharp magnetization reversal and finite coercive field, confirming a large PMA (about $\approx1.2$\,T), in quantitative agreement with previous results~\cite{Manchon2008,Dieny2017}. Upon increasing the Al* thickness to 1.1\,nm, the positive $\mu_0H_K$ drops rapidly and Co magnetization turns in-plane ($\mu_0H_K$ reaching negative value of about $-70$\,mT) as shown by green and black loops in the inset for $t_{\rm Al}*$ = 1.1 and 1.2\,nm, respectively. The anisotropy is found to be negative for $t_{\rm Al}*$ between 1.0 and 1.4\,nm. It is worth to remind that the XPS data confirms the presence of a metallic Al at Co|Al* interface for $t_{\rm Al}* \geq 1.1$\,nm. For $t_{\rm Al}* = 1.4$\,nm, the $\mu_0H_K$ is found to be $+23$~mT, at the verge of the second sign change. At larger $t_{\rm Al}*$, the anisotropy continues to increase until it saturates at about 1.8\,T for $t_{\rm Al}*\geq 2$\,nm (see the square loop for $t_{\rm Al}* = 2.0$\,nm in the inset of Fig.~\ref{Fig4}).

%However, the observation of square hysteresis loop with pronounced PMA for $t_{\rm Al}*$ = 2.0 nm goes beyond the previous understating on the interfacial origin of PMA. Moreover, it is worth highlighting that the observed double inversion in the sign of anisotropy originates from the light element interface, obviating the traditional requisite of spin-orbit coupling (SOC) to induce PMA. }

\medskip

In Fig.~\ref{Fig4}, we also include some points of anisotropy values obtained from two other Pt|Co based metallic systems, namely Pt|Co(0.9 nm)|Cu($t_{\rm Cu}$)|Pt (3~nm) (open blue square) and Pt|Co(0.9 nm)|Al($t_{\rm Al}$)|Pt (3~nm) (open black triangle). Importantly, in these two systems, the presence of a 3-nm thick Pt cap layer prevents completely the light element to become oxidized~\cite{Krishnia2023}. For Pt|Co|Cu($t_{\rm Cu}$)|Pt, we find that the $\mu_0H_K$ remains almost constant with Cu thickness (blue open squares in Fig.~\ref{Fig4}) akin to the symmetric Pt|Co|Pt structure for $t_{\rm Cu}=0$ (0.9\,T). In contrast, the anisotropy in the Pt|Co|Al($t_{\rm Al}$)|Pt series (black open triangles in Fig.~\ref{Fig4}) exhibits an increase with Al thickness, saturating at 1.8\,T for $t_{\rm Al} = 3$\,nm. An interesting observation is that for $t_{\rm Al}$ larger than 2\,nm, the measured anisotropies exactly (within error margins) correspond to the values measured in the Pt|Co|Al* systems, in which a metallic Al film is present. The twofold enhancement in anisotropy found in Pt|Co|Al* (and also in Pt|Co|Al|Pt) compared to Pt|Co|Cu|Pt and symmetric Pt|Co|Pt structure underscores the substantial role of the metallic Co|Al interface on interfacial properties.

\medskip

In order to get some insights, we have performed  \textit{ab initio}-calculations of magnetic anisotropy energy ~\cite{hallal_impurity-induced_2014} and found the same decrease and subsequent enhancement of PMA as observed in the experiments, see Fig.~\ref{Fig5}a. The PMA first decreases as the interfacial oxygen is progressively removed, turning down the hybridization between O-$p_\mathrm{z}$ and Co-$d_\mathrm{z^2}$ interfacial states. Interestingly, the calculations then suggest a strong PMA enhancement in the bulk-like Co layers due to an in-plane tensile strain imposed on Co by the metallic Co|Al interface, as shown in Fig.~\ref{Fig5}a,b. This large PMA enhancement in the ultrathin Co film with an increasing lattice parameter is due to the increased density of states of the in-plane ($d_\mathrm{xy}$,$d_\mathrm{x^2-y^2}$) minority electrons as the Co atoms get further apart. More results are presented in the Supplemental Material~\cite{supmat}, in which the full discussion of the PMA calculations and the second-order perturbation theory analysis are provided, linking the PMA trends to electronic states with a specific orbital character. The general U-shaped trend of the calculated PMA is matching well the experimentally observed one. More in details, from the calculations, it appears that the PMA reduces up to about 1.0\,nm before increasing again, while the experimental curve displays a PMA regain at about 1.4\,nm. This should not be a surprise as the calculations ignore the important, but {\it a priori} constant, role of the bottom Pt|Co interface.

\medskip

%\textcolor{red}{(COM Vincent : we will write again the conclusion after.}

In conclusion, our investigation into the Pt|Co|Al* model spintronic system elucidates the critical role of Co|Al* interface on i-DMI and PMA. First, we establish a clear correlation between the reduction in oxygen concentration at the Co|Al* interface and a subsequent decrease in the net DMI, addressing discrepancies in measured values by different groups. Second, we observe a sign change in anisotropy indicating a transition from out-of-plane to in-plane orientation, at approximately a single aluminum metallic monolayer at the Co interface. Notably, the anisotropy of Co layer undergoes a second sign change, and then saturates with the formation of a thick metallic interface characterized by enhanced PMA due to strong orbital hybridization at Co|Al interface as revealed by XAS and DFT. Our experimental results reveal a new way to tune the interfacial properties with light metals interface, notably the i-DMI and PMA ~\cite{Tristan2024PRA}, for manipulation of next generation chiral spin textures.

\section*{Acknowledgments}
The authors acknowledge Yanis Sassi for fruitful advices and discussions. This work has been supported by a France 2030 government grant managed by the French national research agency (ANR) PEPR SPIN CHIREX (ANR-22-EXSP-0002), SPINMAT (ANR-22-EXSP-0007) and SPINTHEORY (ANR-22-EXSP-0009), by the ANR with TOPO3D (ANR-22-CE92-0082) and by the European Horizon Europe Framework Programme under grant no. 101135729 (SkyANN). This work contains results obtained from the experiments performed at the Ernst Ruska-Centre (ER-C) for Microscopy and Spectroscopy with Electrons at the Forschungszentrum Jülich (FZJ) in Germany. The ER-C beam-time access was provided via the DFG Core Facility Project ER-C E-014. This project has received funding from the European Union’s Horizon 2020 research and innovation program under Grant Agreement No. 800945 (NUMERICS–H2020-MSCA-COFUND-2017).

\section*{References}

\end{document}